\documentclass{emulateapj}

\usepackage{amsmath}

\shorttitle{Beyond $H_0$ and $q_0$}
\shortauthors{Neben and Turner}

\begin{document}

\title{BEYOND $H_0$ AND $\lowercase{q}_0$:  COSMOLOGY IS NO LONGER JUST TWO NUMBERS}

\author{Abraham R. Neben}
\affil{Kavli Institute for Astrophysics and Space Research and Department of Physics and Massachusetts Institute of Technology, Cambridge, MA 02139, USA; abrahamn@mit.edu}
\and
\author{Michael S. Turner}
\affil{Kavli Institute for Cosmological Physics, Departments of Astronomy and Astrophysics and of Physics, The University of Chicago, Chicago, IL 60637-1433, USA}

\begin{abstract}
For decades, $H_0$ and $q_0$ were the quest of cosmology, as they promised to characterize our ``world model'' without reference to a specific cosmological framework. Using Monte Carlo simulations, we show that $q_0$ cannot be directly measured using distance indicators with both accuracy (without offset away from its true value) and precision (small error bar). While $H_0$ \textit{can} be measured with accuracy and precision, to avoid a small bias in its direct measurement (of order 5\%) we demonstrate that the pair $H_0$ and $\Omega_M$ (assuming flatness and $w = -1$) is a better choice of two parameters, even if our world model is not precisely $\Lambda$CDM. We illustrate with analysis of the Constitution set of supernovae and indirectly infer $q_0 = -0.57 \pm -0.04$. Finally, we show that it may be possible to directly determine $q_0$ with both accuracy and precision using the time dependence of redshifts (``redshift drift'').
\end{abstract}

\keywords{cosmological parameters -- methods: numerical -- supernovae: general}

\section{Introduction}
Cosmology has been characterized as the search for two numbers \citep{sandage1970}: $H_0$, the present expansion rate, and $q_0$, the present deceleration parameter.  Few would argue with the statement that cosmology is a much grander enterprise today, and that our ``world model"  is better described by a larger set of physically motivated parameters, including the energy densities of radiation, dark matter, dark energy, and the equation of state of dark energy.

In this paper we show that in fact, the deceleration parameter cannot be directly measured using distance indicators with both accuracy and precision.  That is, $q_0$ cannot be determined with statistical precision (small error bar) without incurring a bias away from the true value. While $H_0$ can be measured with both accuracy and precision, avoiding a slightly biased measurement requires a better choice of parameters than $H_0$ and $q_0$, e.g., $H_0$ and $\Omega_M$.

Sandage introduced $H_0$ and $q_0$ to provide a model-independent, kinematic description of the expansion of the Universe. This description begins with a Taylor series for the cosmic scale factor $R(t)$, 
\begin{equation}
\frac{R(t)}{R_0} = 1 + H_0(t-t_0) - \frac{1}{2}q_0H_0^2(t-t_0)^2 + \cdots
\end{equation}
where $H_0 \equiv {\dot R}_0/R_0$ and $q_0 \equiv -({\ddot R}_0/R_0)/H_0^2$.
Using the definition of redshift, $1 + z \equiv R_0/R$, and luminosity distance, $d_L \equiv  (1 + z)r(z)R_0$, and the above expansion, the observable luminosity distance can be expressed in a Taylor series in redshift
\begin{equation}
\label{eqn:hubble_exp}
H_0 d_L =  z + {1\over 2}(1-q_0)z^2 + {O}(z^3)
\end{equation}
Note too, that no assumption about the validity of general relativity has been made; only that spacetime is isotropic and homogeneous and described by a metric theory.  We note that the next order term---the jerk parameter $j_0$---may be added \citep[e.g.][]{jerk2,jerk1,visser05,weinberg}:

$$\frac{R(t)}{R_0} = 1 + H_0(t-t_0) - {1\over 2}q_0H_0^2(t-t_0)^2 + {1\over 6} j_0 H_0^3(t-t_0)^3 + \cdots$$
$$H_0d_L = z + {1\over 2}(1-q_0)z^2 - {1\over 6} \left(1-q_0-3q_0^2 + j_0 + \frac{K }{H_0^2R_0^2}\right)z^3 +\cdots$$
where $j_0 \equiv (\dddot{R}_0/R_0)/H_0^3$ and $K=0$, $-$1, or 1 for a flat, open, or closed universe, respectively.

The power of the $(H_0,q_0)$ (or quadratic) expansion is that, in principle, measurements of $d_L(z)$ can be used to determine the present expansion rate---arguably the most important number in all of cosmology---and the deceleration parameter.  Moreover, in the simple matter-only cosmology of the time, general relativity, through the Friedmann equations, relates $q_0$ to the physical parameter, $\Omega_0 \equiv \rho_M/\rho_{\rm crit}$, where $\rho_M$ is the present matter density and $\rho_{\rm crit} \equiv 3H_0^2/8\pi G$ is the present critical density:  

$$q_0 = \Omega_0/2$$

Further, in this model, $q_0$ is related to the spatial curvature and destiny of the Universe:  if $q_0$ is greater than $1/2$, $\Omega_0$ is greater than unity and the Universe is positively curved and will ultimately re-collapse; conversely, if $q_0$ is less than 1/2, $\Omega_0$ is less than 1 and the Universe is negatively curved and will expand forever.  The case of $q_0 = 1/2$ is the flat Universe that expands forever at an ever slowing rate.

\begin{figure}[h]
\includegraphics[width=8.5cm]{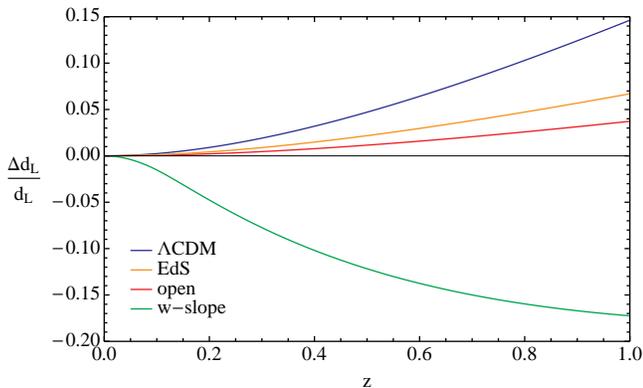}
\caption{Fractional error $\Delta d_L/d_L$ of the quadratic expansion for $d_L$ (Equation \ref{eqn:hubble_exp}) in various cosmologies. Note that $\Delta \mu \approx 2.17 \Delta d_L/d_L$, where $\mu$ is the distance modulus and $\Delta d_L$ and $\Delta \mu$ are the absolute errors of the quadratic  expansions for $d_L$ and $\mu$, respectively.}
\label{fig:hubble_exp}
\end{figure}

This is all well and good; the question we address here is whether these two (or three), model-independent parameters can in fact be measured. The answer is simple:  only $H_0$ can be measured with accuracy and precision.  The explanation is simple as well:  at low redshifts, say $z\lesssim0.2$, where the Taylor expansion is most accurate, poor leverage on $q_0$ (and $j_0$) and peculiar velocities severely limit the precision; at higher redshifts, where the effect of peculiar velocities is negligible and the leverage is greater, the quadratic expansion does not accurately approximate $d_L(z)$ (see Figure \ref{fig:hubble_exp}) and a bias is introduced in the measurement.  One can measure $q_0$ either with precision (use measurements extending to high redshift), or with accuracy (restrict the measurements to low redshift). We also show that including the cubic term in the model-independent expansion---jerk $j_0$---does change this conclusion.

Moreover, while one might be tempted to simply abandon $q_0$ and $j_0$ given these complications, in fact they have found use recently to avoid a bias of order $-5\%$ that otherwise inflicts direct measurements of $H_0$ due to the less accurate linear expansion \citep{riess2009, riess2011}. To better avoid such biases, we explore other two- or three-parameter descriptions of $d_L$.  In particular, the pair $H_0$ and $\Omega_M$ (where flatness and $w = -1$ are assumed) eliminates this small bias in measuring $H_0$ and is better motivated. We use the Constitution compilation of supernovae \citep{constitution} to concretely illustrate of our findings. Finally, we briefly address how $q_0$ might be measured with accuracy and precision by exploiting the very small time evolution of redshifts----of the order of cm s$^{-1}$ decade$^{-1}$---known as redshift drift (Sandage 1962; Loeb 1998).

Our paper is organized as follows. In Section \ref{sec:sims} we detail our Monte Carlo distance indicator surveys and redshift drift surveys.  We present the results of these simulations in Section \ref{sec:results}, including analysis of the Constitution set of supernovae to illustrate our findings.  Section \ref{sec:disc} provides some discussion and conclusions.

\section{Monte Carlo Simulations}
\label{sec:sims}

\subsection{Simulated Distance Indicator Surveys}
\label{sec:mocksurveys}

\begin{figure}[h]
\includegraphics[width=8.5cm]{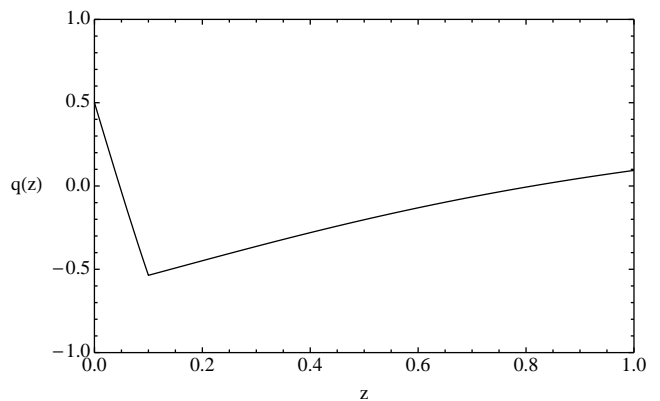}
\caption{Deceleration parameter, $q(z)$, in the w-slope model (see Table~\ref{table:models}).}
\label{fig:wslope_q}
\end{figure}

We use a diverse set of cosmological models (Table~\ref{table:models}) including $\Lambda$CDM, de Sitter (dS), an open model, Einstein--de Sitter (EdS), and a model with rapid dark energy evolution over redshift which we term ``w-slope'' (Figure \ref{fig:wslope_q}). For each, we generate mock distance modulus and redshift $(\mu,z)$  data over various redshift ranges, then study the degree to which $q_0$ and $j_0$ are constrained when these data are analyzed with the quadratic ($H_0,q_0$) or cubic ($H_0,q_0,j_0$) Hubble expansions. We then compare the degree to which $H_0$ is constrained using these expansions to the constraints obtained when the analysis assumes $\Lambda$CDM. In the latter case, $H_0$ and $\Omega_M$ are the free parameters. We also investigate the effects of peculiar velocities and intrinsic luminosity scatter. 

 \begin{table}[h]
 \caption{ \label{table:models}Cosmological Models}
 \begin{tabular}{llllllll}
 \hline\hline
 Model & $\Omega_\text{M}$ & $\Omega_\text{DE}$ & $w_\text{DE}(z)$ & $q_0$\\
 \hline
$\Lambda$CDM & 0.28\tablenotemark{a} & 0.72\tablenotemark{a} & -1 & -0.58\\
de Sitter (dS) & 0 & 1& -1 & -1\\
open & 0.28 & 0 & -- & 0.14\\
Einstein-de Sitter\\\hspace{3mm}(EdS) & 1 & 0 & -- & 0.5\\
w-slope & 0.28 & 0.72 & $\left\{{\small\begin{array}{lr}
       -10z&\text{ for $z<0.1$}\\
       -1 & \text{ for $z>0.1$}
     \end{array}}\right\}$ & 0.5\\ 
     \tableline
 \end{tabular}
 \tablenotetext{1}{\citet{komatsu}.}
 \end{table}
 
We model the distance indicators as a population of imperfect standard candles with an intrinsic Gaussian absolute magnitude scatter $\sigma_\mathrm{int}$. (While we are motivated by and will eventually use Type Ia SNe as the distance indicator, our results are more general.) We explore two possible values of $\sigma_\mathrm{int}$: 0.15 mag, of order today's state of the art for SNe Ia after light curve fitting \citep[e.g.][]{unionPaper, kessler, cfa3Paper, rapetti,conley}; and $0.02$ mag, an optimistic estimate of what distance indicators including type Ia SNe might eventually be able to achieve.  At this precision, peculiar velocity scatter dominates the $\mu$ uncertainties at $z < 0.2$. Note the intrinsic distance modulus scatter is simply related to the luminosity scatter as $\sigma_\mathrm{int} = (d \mu/d L) \sigma_L  = 1.08 (\sigma_L/L)$. 

Each mock measurement of $\mu$ is chosen from a Gaussian distribution of width $\sigma_\mathrm{int}$ centered on the true value of $\mu(z)$, which is given by standard equations \citep[e.g.][]{weinberg}. We then apply a modest peculiar velocity scatter to the redshift data, giving the mock measurement of redshift: $(z+1)(z_\mathrm{pec}+1)-1$, where each $z_\mathrm{pec}$ is chosen from a Gaussian distribution centered at zero with $\sigma_\mathrm{pec}=300$ km s$^{-1}$ \citep[e.g.][]{riess2004,riess2007,rapetti,constitution,lampeitl}. (We note that there is no a priori reason why peculiar velocities should be described by an independent redshift scatter applied to each standard candle, and in general, one would expect  correlations due to bulk flows to prevent statistical uncertainties in fit parameters from decreasing as fast as $1/\sqrt{N}$, where $N$ is the number of data points. However, previous surveys \citep[e.g.][]{lampeitl} and $N$-body simulations \citep{pecVelPaper} have suggested that such correlation is not significant enough to pose a large issue.) We briefly explore the effect of lowering this scatter to the 150 km s$^{-1}$ achieved by Conley et al. (2011) using the local bulk flow model of  Hudson et al. (2004). 

\begin{figure}[h]
\includegraphics[width=8.5cm]{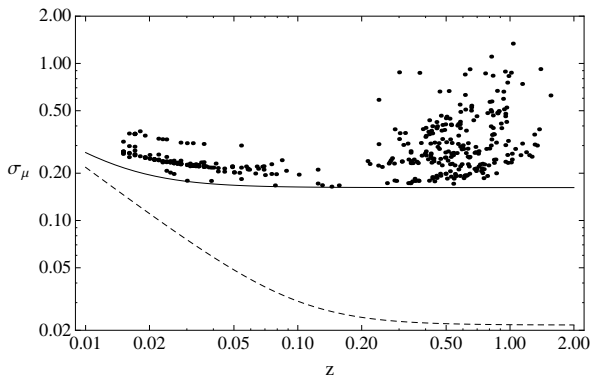}
\caption{Effective distance modulus scatter $\sigma_\mu(z)$ (Equation \ref{eqn:sigma_mu}) used in our simulations with $\sigma_\mathrm{int}=0.15$ (solid) and $\sigma_\mathrm{int}=0.02$ (dashed), as well as the published Constitution set uncertainties with $\sigma_\text{int}=0.15$ added in quadrature.}
\label{fig:variances_plot}
\end{figure}

In parameter estimations, this peculiar velocity scatter can be incorporated as an additional uncertainty in $\mu$ added in quadrature to the intrinsic scatter: $(d \mu/d z) \sigma_\mathrm{pec}=2.17 \sigma_\mathrm{pec}/z$, approximating $\mu(z)$ with the linear  expansion for $d_L$. While others have assumed the empty universe model here \citep[e.g.][]{kessler, lampeitl}, essentially taking the quadratic  expansion with $q_0=0$, we found the difference on our parameter estimates to be negligible. Figure~\ref{fig:variances_plot} shows the effective $\mu$ scatter $\sigma_\mu(z)$,

\begin{equation}
\label{eqn:sigma_mu}
\sigma_\mu(z) = \sqrt{\sigma_\mathrm{int}^2 + (2.17 \sigma_\text{pec}/z)^2}
\end{equation}
as well as the $\mu$ uncertainties in the Constitution data set \citep{constitution} with the intrinsic scatter $\sigma_\mathrm{int}=0.15$ mag added in quadrature.

For a given cosmological model, redshift range, and intrinsic standard candle scatter, we simulate 10,000 surveys, each with 500 mock objects uniformly distributed over redshift.\footnote{For simplicity and to illustrate our basic results, we use a uniform distribution in redshift.  In principle one could try to optimize the redshift distribution and of course in practice, observational considerations influence what can actually be done. \citep{sneSurveyBiases}} We use a minimum redshift of 0.015 \footnote{This is the smallest redshift present in the Constitution set.} and explore several values of the maximum redshift between 0.05 and 1.5 chosen to illustrate the dependence of accuracy and precision in parameter estimates on redshift. Redshifts smaller that 0.015 are unhelpful for measuring $q_0$ due to severe peculiar velocity uncertainties (see Figure \ref{fig:variances_plot}) and the very weak leverage on that parameter at such low $z$. To eliminate simulation biases due to mock redshifts scattering {\it out} of the redshift range $(0.015,z_\mathrm{max})$, but not scattering {\it into} it, we generate more than 500 data points over a slightly larger redshift range, then keep only 500 data points in the desired range.

We then explore biases in precision measurements of $H_0$ due to the inaccuracy of the $d_L$ expansion. To avoid any systematics present in high-$z$ supernova data (such as evolution or dust effects), typically only nearby supernovae are used to measure $H_0$. For these simulations, we take $z_\mathrm{max}=0.1$, which can be identified as the cutoff redshift of the low-$z$ segment of the Constitution set (Figure \ref{fig:variances_plot}). 

\subsection{Parameter Estimation from Simulated Distance Indicator Surveys}

For each mock survey, we estimate model parameters by minimizing a $\chi^2$ statistic. For instance, to estimate $H_0$ and $q_0$ using the quadratic  expansion, we minimize:

\begin{equation}
\label{eqn:log_likelihood}
\chi^2(H_0,q_0)= \sum_{i=1}^{500}\left( \frac{\mu_i - \mu_\mathrm{exp}(H_0,q_0; z_i)}{\sigma_\mu(z_i)} \right)^2
\end{equation}

Then for each cosmological model, redshift range, and intrinsic $\mu$ scatter, we plot 2D contours in $(H_0,q_0)$ parameter space with 68\% and 95\% probability content generated from the 10,000 parameter estimations. We treat the 10,000 ($H_0$,$q_0$) pairs as samples from a 2D Gaussian, compute the covariance matrix, then plot contours of that Gaussian. We have verified that the error in assuming the distribution is Gaussian is not significant for our purposes. Each contour plot shows the probability distribution of ($H_0$,$q_0$) pairs in parameter space given the intrinsic $\mu$ scatter and peculiar velocity scatter the data are drawn from, showing any bias in and correlation between the parameter estimates. The extents of the contours may be taken as error ellipses for a single measurement. 

\subsection{Comparison with Real Data}

To compare our results with real data, we use the Constitution set (Table 1 in \citet{constitution}), which combines the low redshift ($z\lesssim0.08$) CfA3 sample of \citet{cfa3Paper} with the Union sample of \citet{unionPaper}, processed with the SALT light-curve fitter of \citet{saltPaper}. We use the published $(z,\mu,\Delta\mu)$ data, where $\Delta\mu$ includes some systematic uncertainties as discussed by \citet{constitution}. Note that we use this data set only to constrain $q_0$ and $j_0$, not to make precision $H_0$ measurements.

When analyzing the Constitution set of supernovae, we pick several illustrative maximum redshifts to compare results with simulations. Unlike with our mock data sets, the number of objects is not the same for each redshift range. The numbers of SNe up to $z_\mathrm{max}=\{0.1,0.3,0.5,1.5\}$ are \{141, 164, 248, 397\}. Following standard practice \citep[e.g.][]{lampeitl,kessler,rapetti}, we add the intrinsic standard candle scatter (0.15 mag) in quadrature to the published $\mu$ uncertainties.

\subsection{Simulated Redshift Drift Surveys}
\label{sec:dzdtsurveys}

The redshift of an object within the Hubble flow varies with time due to the expansion of the Universe \citep{dzdtSandage,dzdtLoeb}:
$$ \frac{dz}{dt} = (1+z) H_0 - H(z) \longrightarrow -zq_0H_0 + O(z^2) $$
where the $(H_0,q_0)$  expansion has been used to obtain the final expression. The estimate for $q_0$ using this method, assuming $H_0$ has been determined by other means, is given by

\begin{equation}
\label{eqn:q0est}
q_0\approx-\frac{1}{H_0 z}\frac{dz}{dt}
\end{equation}Note, that unlike in the quadratic Hubble expansion, both $H_0$ and $q_0$ appear together at lowest order (linear in $z$) here. 

We assume $dz/dt$ is observed for $N$ distance indicators spaced uniformly over redshift from $z_\text{min}=0.015$ to $z_\text{max}$, exploring $N=500$, 5000 and various values of $z_\text{max}$. We assume a peculiar acceleration scatter with mean of zero and standard deviation of 200 km$^{-1}$ s$^{-1}$ (50 Myr)$^{-1}$ = 4 cm$^{-1}$ s$^{-1}$ decade$^{-1}$, estimating this numerical value from the orbital period and circular velocity of our solar system around the Milky Way. This is of order the cosmological signal in $\Lambda$CDM \citep[e.g.][]{amendola,quercellini2012} which is smaller than 6 cm$^{-1}$ s$^{-1}$ decade$^{-1}$ for $z<2$; therefore, as long as the instrumental uncertainties on these measurements are smaller than $\sim50$\%, they will be subdominant to the peculiar accelerations, and we thus neglect them. We perform 10,000 realizations for each set of parameters, then take the standard deviation of the resultant $q_0$ distribution as the statistical uncertainty of a single measurement, and the deviation of the mean away from the true value as the bias.

\section{Results}
\label{sec:results}

\subsection{Measuring $q_0$}
\label{sec:measuringq0}

\begin{figure*}[t!]
\includegraphics[width=18cm]{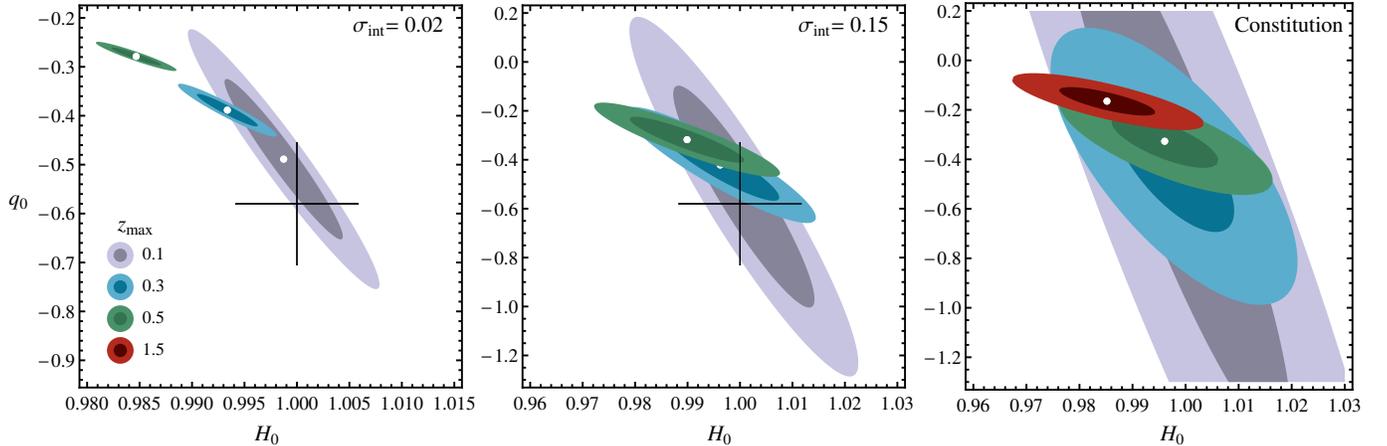}
\caption{Contours in $(q_0,j_0)$ space from $\Lambda$CDM simulations analyzed with quadratic expansion (left and center), and contours from analysis of Constitution data (right). The crosshairs indicate the true values $(H_0,q_0)=(1,-0.58)$ in $\Lambda$CDM, where $H_0$ is in units of its true value. The improvement in precision at the expense of worsening accuracy (growing bias) is evident both in the simulations and in the Constitution data as $z_\mathrm{max}$ increases. Constitution $H_0$ values are in units of 65 km s$^{-1}$ Mpc$^{-1}$, as used in \citep{constitution}. Note the expanded axes of the left-most plot.}
\label{fig:cc}
\end{figure*}

\begin{figure}[h]
\includegraphics[width=8.5cm]{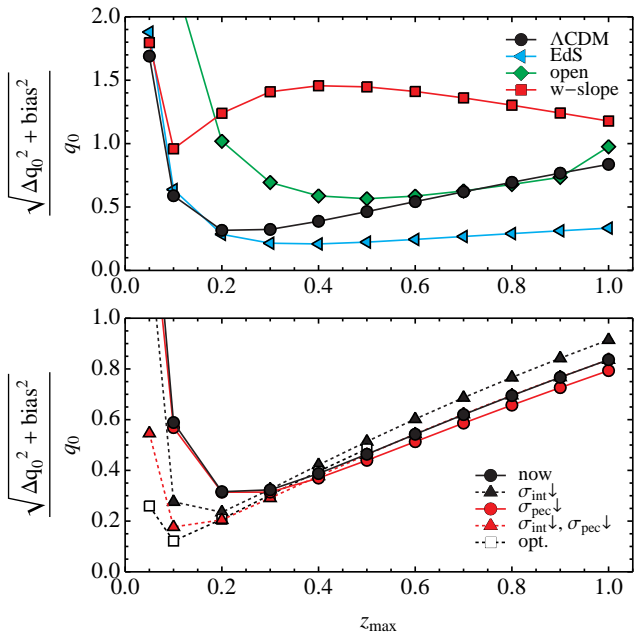}
\caption{Quadrature sum of statistical uncertainty and bias in $q_0$ estimates from simulated distance indicator surveys fit to the quadratic expansion. The top panel shows various cosmological models all with $\sigma_\text{int}=0.15$. The bottom panel shows $\Lambda$CDM with various survey improvements such as lowering $\sigma_\text{int}$ to 0.02 and lowering $\sigma_\text{pec}$ to 150 km s$^{-1}$. The ``opt'' simulations represent an optimistic survey scenario with substantial improvements (N=5000, $\sigma_\text{int}=0.02$, $z_\text{min}=0.01$, and $\sigma_\mathrm{pec}=150$ km sec$^{-1}$) over our nominal assumptions. In all cases, lack of leverage on $q_0$ combined with peculiar velocities drastically limits precision at low redshift (statistical scatter) while the inexactness of the quadratic expansion limits accuracy at higher redshift (growing bias).}
\label{fig:errorfloor}
\end{figure}

Figure \ref{fig:cc} shows $(H_0,q_0)$ contour plots from analysis of 10,000 $\Lambda$CDM simulations with the quadratic  expansion, as well as contours from analysis of Constitution data. The trend with $z_{\rm max}$ (and our basic result) is clearly illustrated:  increasing $z_{\rm max}$ improves precision of the $q_0$ estimate only at the expense of a growing bias away from the true value. This basic trend is seen for all the cosmological models (except deSitter, see below), and is simply illustrated in the top panel of Figure \ref{fig:errorfloor}, which shows the quadrature sum of statistical uncertainty and bias for all the models as a function of $z_\text{max}$ (all with $\sigma_\mathrm{int}=0.15$). The ``sweet spot'' where both accuracy and precision are decent varies between $z_\text{max}\sim0.1- 0.5$ depending on the model. For $\Lambda$CDM, a minimum of 32\% is reached at $z_\text{max}=0.25$ with $\Delta z_\text{max}=0.18$ (within 20\% of the minimum).

The bottom panel of Figure \ref{fig:errorfloor} shows the effects of varying some of our data quality assumptions. Lowering $\sigma_\mathrm{pec}$ to 150 km/sec has a negligible effect on $q_0$ precision unless $\sigma_\mathrm{int}$ is lowered significantly as well. Increasing survey size and lowering $z_\mathrm{min}$ help further, giving a minimum of 13\% at $z_\text{max}=0.10$ with $\Delta z_\text{max}=0.06$ (within 20\% of the minimum). None of these improvements is a `silver bullet'; all are required to tackle the large error bars produced in these $q_0$ estimates due to the very weak statistical leverage on that parameter in the quadratic expansion.

Returning to Figure \ref{fig:cc}, we more carefully study shrinking the intrinsic $\mu$ scatter in the left panel. Lowering $\sigma_\mathrm{int}$ to 0.02 mag, the regime where parameter uncertainties in the $z_\text{max}=0.1$ simulations are dominated by peculiar velocity scatter, reduces the $q_0$ uncertainties by a factor of $\sim$4, though the biases worsen slightly due to the altered distribution of uncertainties as a function of $z$ (high $z$ data now weighted much more relative to low $z$ data, see Figure  \ref{fig:variances_plot}). We note here that while our simulations with $\sigma_\mathrm{int}=0.15$ mag are intended to roughly approximate present day SNe Ia surveys, the Constitution data set contains many objects with much larger uncertainties and the set is distributed non-uniformly over redshift (Figure \ref{fig:variances_plot}). Thus, while the Constitution $(H_0,q_0)$ contours exhibit the same trends as those in the $\sigma_\mathrm{int}=0.15$ simulation plot, the Constitution uncertainties are larger. 

\begin{figure*}[h]
\centering
\includegraphics[height=23.2cm]{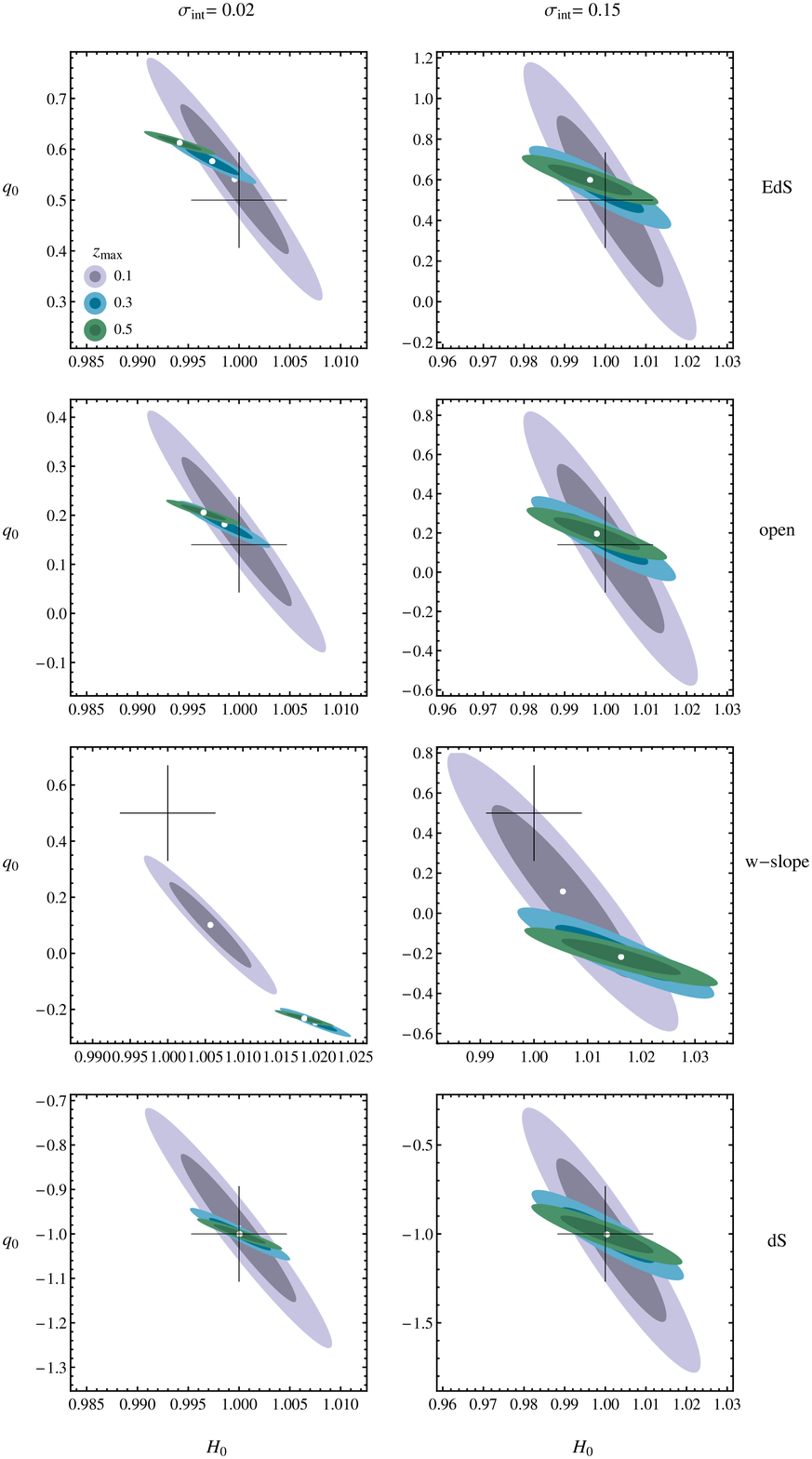}
\caption{Contours in $(H_0,q_0)$ space from simulated distance indicator surveys in various cosmological models (see Table~\ref{table:models}) analyzed with quadratic expansion. The crosshairs indicate the true $(H_0,q_0)$ in each model, where $H_0$ is in units of its true value. The improvement in precision at the expense of worsening accuracy (growing bias) is visible in all models except the de Sitter model, in which the quadratic expansion is exact. Estimates of $q_0$ in the w-slope model are especially poor owing to its rapid evolution over redshift (see Figure \ref{fig:wslope_q}). Note the expanded axes in the left column.}
\label{fig:other_models}
\end{figure*}

Similar contour plots for other cosmological models appear in Figure \ref{fig:other_models}. Again we observe the expected trade-off between accuracy at low $z$ and precision at high $z$. The open and EdS models are somewhat better approximated by their quadratic  expansions than is $\Lambda$CDM (see Figure \ref{fig:hubble_exp}), and thus, they admit slightly more accurate $q_0$ estimates. Estimates of $q_0$ in the w-slope model are especially poor owing to its rapid evolution over redshift (see Figure \ref{fig:wslope_q}). The de Sitter model is worth mentioning because it provides a check on the reliability of the Monte Carlo simulations. In that model, the quadratic  expansion is exact, a fact reflected in our parameter estimates by the absence of any biases.

\begin{figure*}[t!]
\includegraphics[width=18cm]{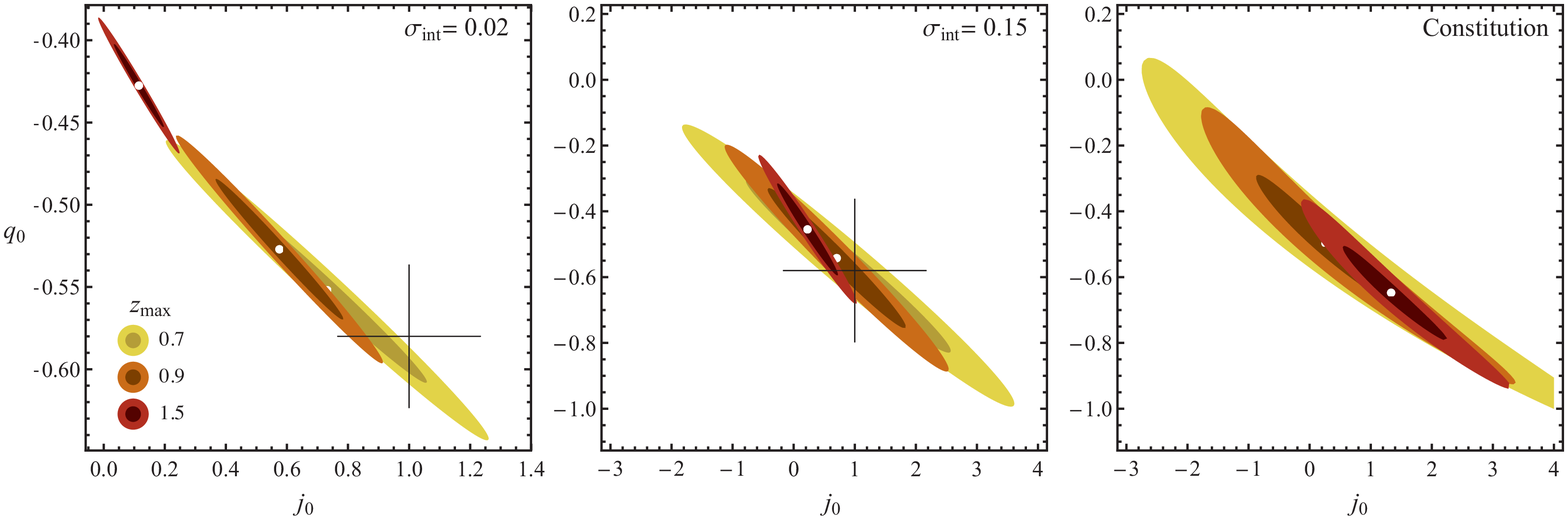}
\caption{Contours in $(q_0,j_0)$ space from simulated distance indicator surveys in $\Lambda$CDM analyzed with cubic expansion (after marginalizing over $H_0$) (left and center), as well as contours from analysis of Constitution data (right). The crosshairs indicate the true $(j_0,q_0)=(1,-0.58)$ in $\Lambda$CDM. As the bias in the parameter estimates is reduced (relative to estimates using quadratic expansion), we show the results of simulations over larger redshift ranges where the compromise between accuracy and precision is clearer. Note the expanded axes of the left-most plot.}
\label{fig:cc_jerk}
\end{figure*}

Figure \ref{fig:cc_jerk} shows the results of including the cubic term when analyzing data from the $\Lambda$CDM model  in the form of $(j_0,q_0)$ contour plots after marginalizing over $H_0$. The overall trend is the same as before: increasing $z_{\rm max}$ improves precision at the expense of a growing bias.  Relative to the quadratic expansion, the bias is reduced and we thus use $z_\mathrm{max} = 0.7, 0.9$, and 1.5. Reducing $\sigma_\mathrm{int}$ to 0.02 mag helps somewhat, but as before worsens the biases slightly. Using the entire Constitution set ($z_\mathrm{max}=1.5$), we estimate $q_0=-0.64\pm0.14$, within error bars of our simulation result of $-0.45\pm0.09$, with slight differences possibly due to the non-uniform distribution of redshifts and non-uniform errors in the Constitution set. 

\begin{figure}[h]
\includegraphics[width=8.5cm]{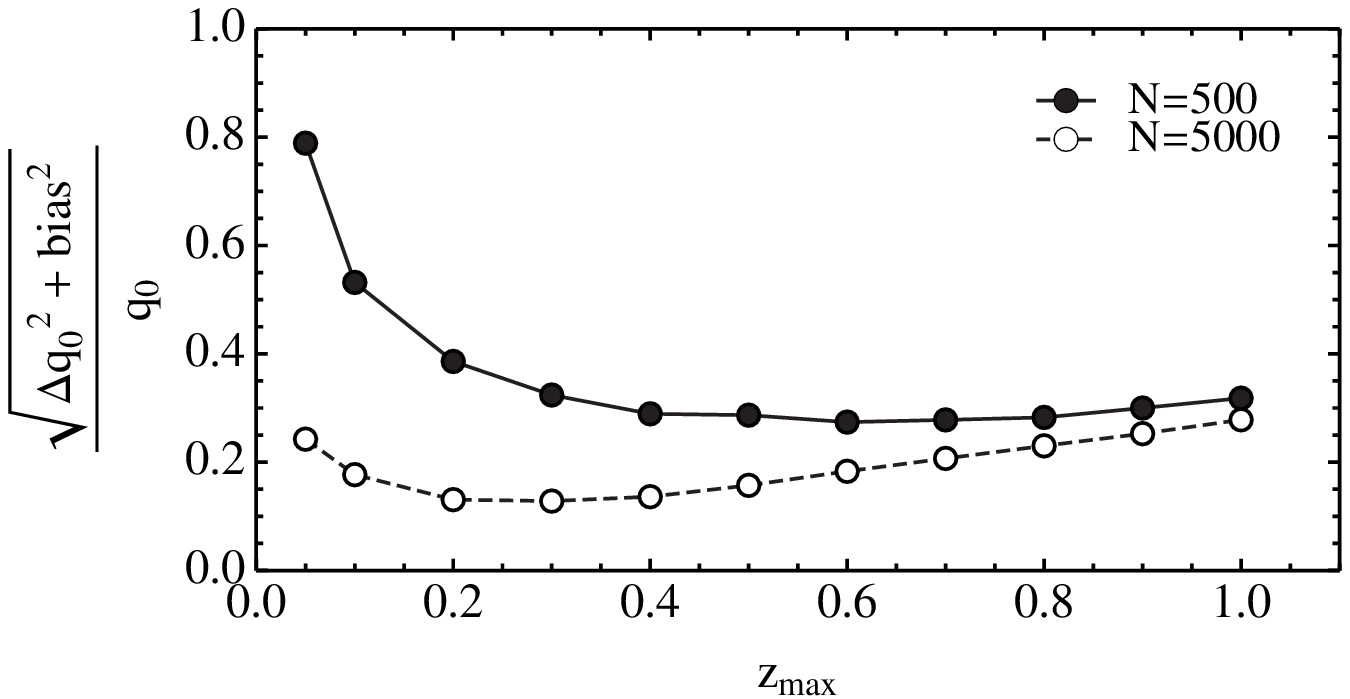}
\caption{Quadrature sum of statistical uncertainty and bias in $q_0$ estimates from simulated redshift drift surveys fit to Equation \ref{eqn:q0est}, with $\Lambda$CDM as the underlying cosmology. Compared to our distance indicator simulations, a somewhat broader redshift `sweet spot' over which both accuracy and precision are decent is observed here (cf. Figure \ref{fig:errorfloor}). The precision is also comparatively improved at small $z$. }
\label{fig:errorfloordzdt}
\end{figure}

Turning to our simulated redshift drift surveys, Figure \ref{fig:errorfloordzdt} shows the quadrature sum of statistical uncertainty and bias in the $q_0$ estimate using this method, with $\Lambda$CDM as the underlying cosmology. Again we find that precision worsens at small $z$, this time because the $1/z$ factor in Equation \ref{eqn:q0est} magnifies peculiar acceleration uncertainties as $z\to0$. Still, the redshift `sweet spot' is somewhat broader than in our distance indicator simulations (cf. Figure \ref{fig:errorfloor}), and precision at small $z$ is greatly improved. A minimum of 29\% is reached at $z_\text{max}=0.65$ with $\Delta z_\text{max}=0.7$ (within 20\% of the minimum) for $N=500$ objects, and 12\% at $z_\text{max}=0.25$ with $\Delta z_\text{max}=0.3$ (within 20\% of the minimum) for $N=5000$.

\begin{figure*}[t]
\centering
\includegraphics[height=5.75cm]{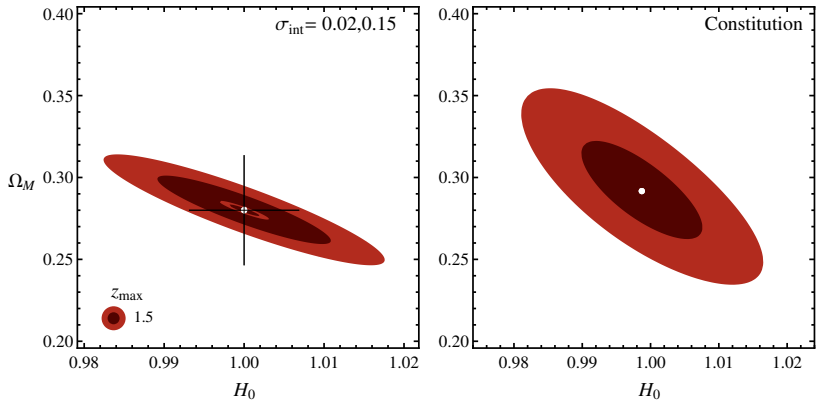}
\caption{$(H_0,\Omega_M)$ contours from $\Lambda$CDM simulations analyzed with the standard $(H_0,\Omega_M)$ parameterization (assuming flatness and $w=-1$), plotted with contours from same analysis on Constitution data. The crosshairs indicate the true simulation values $(H_0,\Omega_M)=(1,0.28)$. Simulation $H_0$ values in units of true $H_0$, Constitution $H_0$ values in units of 65 km sec$^{-1}$ Mpc$^{-1}$, as used in \citep{constitution}. Note the absence of any large bias in either parameter.\vspace{7mm}}
\label{fig:cc_omega_m}
\end{figure*}

As a final check on the reliability of our simulations, Figure \ref{fig:cc_omega_m} shows the analysis of $\Lambda$CDM simulations and of Constitution data with the ($H_0$,$\Omega_M$) parameterization, showing there should be no large biases in determining $H_0$ and $\Omega_M$.  Using the Constitution data and the relation $q_0 = 3\Omega_M/2 - 1$ (assuming $\Lambda$CDM, which is parameterized by $H_0$ and $\Omega_M$), we infer $q_0  = -0.57 \pm 0.04$, with better accuracy (closer to true value) \textit{and} precision (smaller error bar) than estimates derived from the quadratic or cubic expansions. Here we have used the published Constitution uncertainties, meaning this error bar even includes some systematic uncertainties (see Section 2.3).

\subsection{Measuring $H_0$}

\begin{figure}[h]
\includegraphics[width=8.7cm]{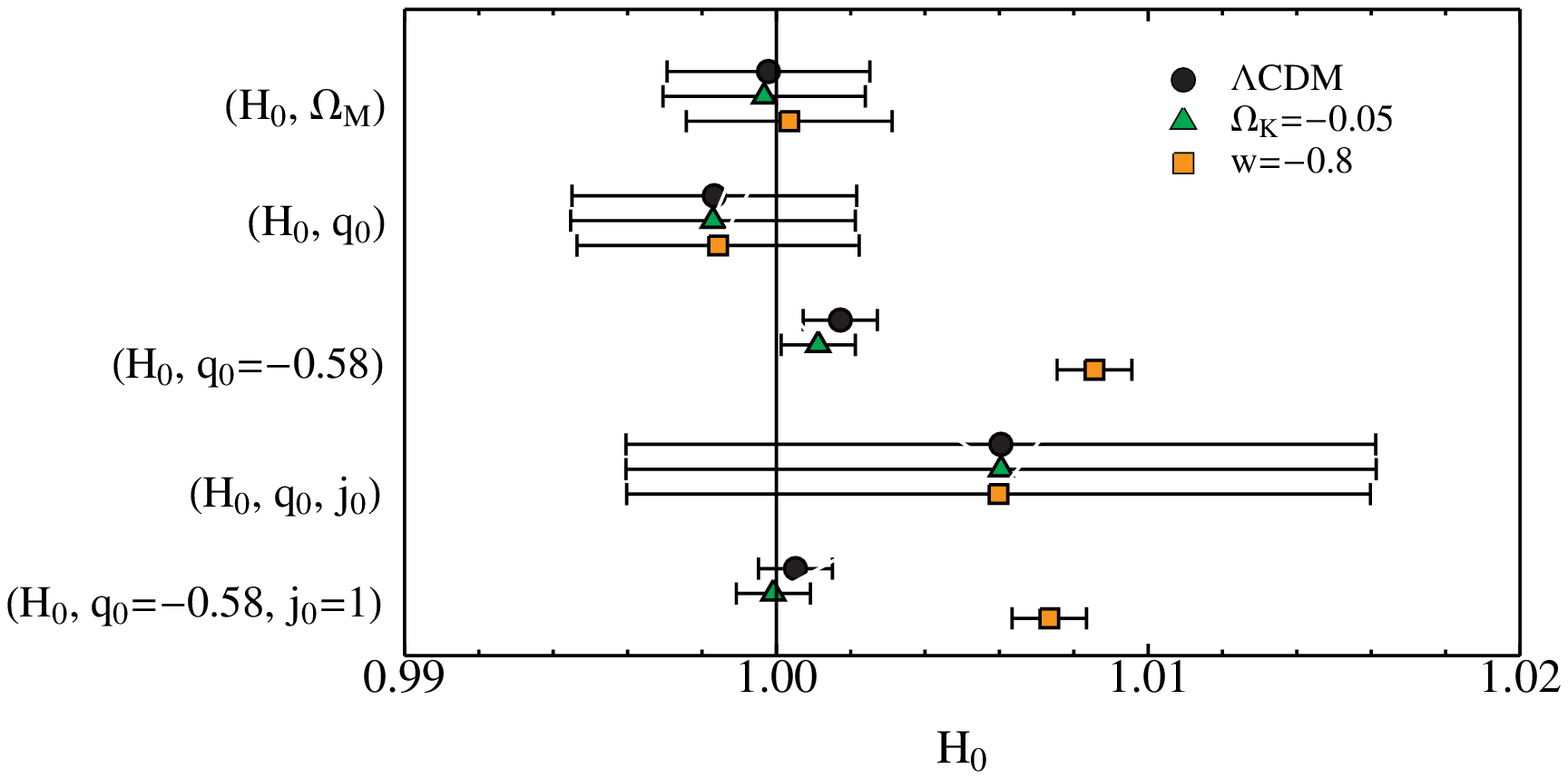}
\caption{$H_0$ estimates in $\Lambda$CDM and cosmologies close to $\Lambda$CDM obtained by fitting simulated distance indicator surveys ($\sigma_\mathrm{int}=0.02$ and $z_\mathrm{max}=0.1$) to five different parameterizations. With $\sigma_\mathrm{int}=0.15$, $H_0$ uncertainties are a factor of 2--3 larger and biases are comparable.}
\label{fig:H0_bias}
\end{figure}

Next, we explore the small biases that creep into precision $H_0$ measurements due to the inaccuracy of the quadratic and cubic expansions. We simulate distance indicator surveys with $z_\mathrm{max}=0.1$ (as discussed in Sec.~\ref{sec:mocksurveys}) in $\Lambda$CDM and in two models close to $\Lambda$CDM but still consistent with current data ($\Lambda$CDM may not be the exact world model):  a model with $w=-0.8$ that is otherwise identical to $\Lambda$CDM; and a slightly closed model with $w=-1$, $\Omega_M=0.30$, and $\Omega_\text{DE}=0.75$ ($\Omega_K=-0.05$).  We then estimate $H_0$ by fitting to:

\begin{enumerate}
\item ($H_0,\Omega_M$) parameterization of $\Lambda$CDM (assuming flatness and $w=-1$);
\item quadratic  expansion, marginalizing over $q_0$;
\item quadratic  expansion, fixing $q_0=-0.58$;
\item cubic expansion, marginalizing over $q_0$ and $j_0$;
\item cubic expansion, fixing $q_0=-0.58$ and $j_0=1$.
\end{enumerate}

The results are shown in Figure \ref{fig:H0_bias}. We note first that using the linear  expansion for $d_L$ yields a bias in $H_0$ (not shown) of $-4\%$ for $\sigma_\mathrm{int}=0.15$, and $-5\%$ for $\sigma_\mathrm{int}=0.02$, in good agreement with with the $-3\%$ bias identified by \citet{riess2009} using real data. Using the quadratic or cubic  expansion and marginalizing over the other parameters reduces these biases to order $0.5\%$, and assuming values for those parameters reduces them further (except for the $w=-0.8$ model). 

Even if our world model is not $\Lambda$CDM, the $(H_0,\Omega_M)$ parameterization is the most robust, with biases of order $0.01\%$. With $\sigma_\mathrm{int}=0.02$, peculiar velocities dominate over $\mu$ uncertainties at these redshifts, and provide the limiting uncertainty in $H_0$. They are also the source of the lingering $\sim0.01\%$ biases, as the peculiar velocity uncertainties are estimated using \textit{measured} $z$, not actual $z$, so objects scattering to lower (higher) $z$ are under- (over-) weighted in the likelihood analysis.

\section{Discussion}
\label{sec:disc}
The quest to measure $H_0$ and $q_0$ and determine our world model drove cosmology for almost three decades.  The Hubble constant has now  been directly measured to around 5\% precision \citep{riess2011}, inferred from cosmic microwave background (CMB) anisotropy and other cosmological measurements to almost 1\% \citep{komatsu}, and there are aspirations to improve that to less than a percent \citep[e.g.][]{riess2011,hubbleConstConf,WeinbergCosmicAccel,ReidJCAP,FreedmanMadore,Seki}.  $H_0$ remains arguably the most important single number in cosmology, as it sets the age and size of the Universe and underpins many other measurements.

On the other hand, not only was the minus sign in the definition of $q_0$ unfounded, but $q_0$ cannot actually be measured with precision and accuracy using distance indicators!  The reasons are simple:  at low redshifts, where the $d_L$ expansion is accurate, the combination of peculiar velocities and the small change in $d_L$ for different values of $q_0$ (lack of leverage)  makes $q_0$ impossible to measure with any precision.  At higher redshift, where there is leverage and precision is possible, the inaccuracy of the $(H_0,q_0)$ expansion strongly biases estimates of $q_0$. There is no ``sweet spot'' at intermediate redshift that allows both precision and accuracy regardless of the world model (see Figure \ref{fig:errorfloor}).

Moreover, the mere usage of $q_0$ as the second parameter in precision measurements of $H_0$ leads to a non-negligible bias.  As we and others have shown \citep{riess2009}, fitting to a linear Hubble law introduces a bias of order $-5\%$.  One solution, used by \citet{riess2009,riess2011}, is to use the $(H_0,q_0,j_0)$ expansion and by fiat impose the `correct' values of $q_0$ and $j_0$ obtained with $z\sim1$ data. This approach has several drawbacks:

\begin{enumerate}
\item Neglecting the uncertainties on $q_0$ and $j_0$ priors \textit{underestimates} the real uncertainty in $H_0$, providing a false sense of precision. In our simulations, going from \textit{fixing} those values to \textit{freeing} them increases the $H_0$ uncertainty by a factor of 10 (see Figure \ref{fig:H0_bias}), an amount that will be important in future precision attempts to further constrain $H_0$ using this method.
\item $H_0$ estimates using this method are no longer independent of high-$z$ SNe and their potential systematics.
\item This method is not robust to small changes in $w$, acquiring a $\sim1\%$ bias if our world model actually has $w=-0.8$ (see Figure \ref{fig:H0_bias}), within the error bars of recent measurements \citep{constitution,komatsu,sullivan2011}.
\item A generic problem with the cubic expansion is that there are more parameters than there are coefficients to fit for \citep{jerk2,visser05}. Simply neglecting the curvature term removes this problem, however it should be noted that purported estimates of $j_0$ using this method are actually estimates of $j_0+ K/(H_0 R_0)^2$
\end{enumerate}
  
Instead and what we think is better, one can use the more physical two-parameter description of our cosmological model, $(H_0,\Omega_M)$.  If our world model is indeed $\Lambda$CDM, then these are the only two parameters needed to describe $d_L(z)$.  In this case, there is essentially no bias in determining the Hubble constant and only peculiar velocities limit the precision (see Figure \ref{fig:H0_bias}), and even they may be mitigable to some extent using bulk flow models \citep{hudson,conley}. The same figure shows that even if $\Lambda$CDM does not exactly describe our world model, the biases are still very small.

Additional parameters can be added (and may be needed)  to completely characterize $d_L$: for example, $w$ or $w_0$ and $w_a$, and $\Omega_k$, but given the closeness of our world model to $\Lambda$CDM, $(H_0,\Omega_M)$ is a set of two that affords accuracy and precision.  Further, we note that within $\Lambda$CDM additional parameters are needed to characterize CMB anisotropy and other dynamical aspects of the Universe.  The standard six-parameter set is $\Omega_Bh^2$ (baryon density), $\Omega_Mh^2$ (total matter density), $\Omega_{DE}$, $n_S$ (power-law index of density perturbations), $\tau$ (optical depth to last scattering), and $\Delta^2$ (overall amplitude of the spectrum of inhomogeneity);  \citep[e.g.][]{komatsu}, and even more parameters can be added.  Cosmology today is much richer than the two numbers that Sandage used to characterize our world model and the goal of cosmology.

All this being said, there is a certain elegance and utility to the two parameters $H_0$ and $q_0$, even if one of them is difficult to directly measure.  For example, both can be defined independently of general relativity and its Friedmann equations (all that is required is the assumption of isotropy and homogeneity and a metric theory; \citet{riessturner,shapiroturner}).  Together, they characterize the most basic features of the expansion---rate of expansion and whether the expansion is slowing down or speeding up.  Within the parameterizations mentioned above, $q_0$ can be inferred; for example, with the two parameters $\Omega_M$ and $H_0$:
$$q_0 = {3\over 2}\Omega_M - 1 = -0.57 \pm 0.04$$
where the numerical value has been determined from the Constitution data set (assuming $\Lambda$CDM), as discussed in Sec. 3.1, and is in agreement with the true value of $-0.58$ for $\Lambda$CDM with $\Omega_M = 0.28$.

Finally, we have demonstrated with simulations that surveys of redshift drift have the potential to better directly determine $q_0$ with accuracy and precision (see Fig. 8).  This method is more powerful because $dz/dt$ depends upon the product of $H_0$ and $q_0$ and is linear in redshift, allowing good constraints over somewhat wider redshift ranges and improved precision near $z=0$. Of course, this is a very challenging measurement---the effect is on the order of a few cm sec$^{-1}$ decade$^{-1}$ (of order peculiar accelerations for $z < 2$)---and a reach goal for the next generation of extremely large optical telescopes \citep[e.g.][]{dzdtBalbi,corasaniti,amendola,dzdtLiske,quercellini2012}.

Cosmology has changed dramatically since Sandage characterized it as the quest for two numbers.  It has become a precision science characterized by a larger set of more physically motivated numbers.  While, $q_0$ is not actually measurable using luminosity distances and a hindrance to accurately measuring $H_0$, it is nonetheless of interesting parameter in cosmology today---and still not directly measured.

\acknowledgments

This work was supported in part by the National Science Foundation and the Department of Energy. A.R.N. acknowledges support from the MIT Bruno Rossi Graduate Fellowship in Astrophysics. We thank the referee for helpful comments on our manuscript.

\end{document}